\documentclass[12pt,aps,nofootinbib,preprintnumbers,groupedaddress,letterpaper]{article}

\usepackage[dvips]{color}
\usepackage[normalem]{ulem}
\usepackage{amsmath}
\usepackage{enumerate}
\usepackage{amsfonts}
\usepackage{epsfig}
\usepackage{yfonts}

\usepackage{graphicx}
\usepackage{bm}
\usepackage{subfigure}

\newcommand\comment[1]{}

\newcommand\poincare{Poincar\' e }

\newcommand\ov{\over }

\def\le{\left}
\def\ri{\right}

\def\({\left(}
\def\){\right)}
\def\<{\langle}
\def\>{\rangle}

\newcommand\half{{\ensuremath{\frac{1}{2}}}}
\newcommand\p{\ensuremath{\partial}}

\newcommand\field[1]{{\ensuremath{\mathbb{{#1}}}}}

\newcommand{\CC}{\field{C}}

\newcommand{\RR}{\field{R}}

\newcommand{\be}{\begin{equation}}
\newcommand{\ee}{\end{equation}}
\newcommand{\bea}{\begin{eqnarray}}
\newcommand{\eea}{\end{eqnarray}}
\newcommand{\bwt}{\begin{widetext}}
\newcommand{\ewt}{\end{widetext}}

\newcommand{\bi}{\begin{itemize}}
\newcommand{\ei}{\end{itemize}}
\newcommand{\ben}{\begin{enumerate}}
\newcommand{\een}{\end{enumerate}}
\newcommand{\bca}{\begin{cases}}
\newcommand{\eca}{\end{cases}}
\newcommand{\bln}{\begin{align}}
\newcommand{\eln}{\end{align}}
\newcommand{\bst}{\begin{split}}
\newcommand{\est}{\end{split}}

\begin{document}

\begin{titlepage}

\begin{flushright}
QMUL-PH-19-27\\
\end{flushright}

\vspace{5mm}

  \begin{center}

\centerline{\Large \bf {Celestial fields on the string}}
\vskip 0.3cm
\centerline{\Large \bf {and the Schwarzian action}}

\bigskip
\bigskip

{\bf David Vegh}

\bigskip

\small{
{ \it  Centre for Research in String Theory, School of Physics and Astronomy \\
Queen Mary University of London, 327 Mile End Road, London E1 4NS, UK}}

\medskip

{\it email:} \texttt{d.vegh@qmul.ac.uk}

\medskip

{\it \today}

\bigskip


\begin{abstract}

\hskip -0.27cm This paper describes the motion of a classical Nambu-Goto string in three-dimensional anti-de Sitter spacetime in terms of two `celestial' fields on the worldsheet. The fields correspond to retarded and advanced boundary times at which null rays emanating from the string reach the boundary.
The formalism allows for a simple derivation of the Schwarzian action for near-AdS$_2$ embeddings.

\end{abstract}

\end{center}

\end{titlepage}

\tableofcontents
\clearpage

\section{Introduction}

This paper is concerned with classical Nambu-Goto strings in three-dimensional anti-de Sitter (AdS) spacetime. The system is interesting for many reasons. In  general curved spacetimes, the worldsheet  conformal field theory is too complicated to solve. Maximally symmetric spacetimes provide an interesting middle ground between such theories and the free theory in flat target space.
Perturbed long strings in AdS spacetime are described by non-linear equations and thus they provide a simple laboratory for studying non-linear phenomena such as wave turbulence and energy cascades \cite{Ishii:2015wua, Vegh:2018dda}.

Strings in AdS naturally show up in the context of the celebrated gauge/gravity duality \cite{Maldacena:1997re, Gubser:1998bc, Witten:1998qj}. According to the correspondence, a long string on the gravity side ending on the boundary is nothing but the dual of a flux tube stretching between external quarks in the boundary field theory. Long strings have been studied to  calculate, for instance, the drag force on an external quark moving in a thermal plasma \cite{Herzog:2006gh, Liu:2006ug, Gubser:2006bz}.

In a remarkable paper, Maldacena, Shenker, and Stanford discovered a bound on the rate of growth of chaos in thermal quantum systems \cite{Maldacena:2015waa}. This universal ``chaos bound'' on the Lyapunov exponent is $\lambda_L \leq {2\pi T}$ where $T$ is the temperature.
In holographic theories with classical gravity duals, black holes saturate this bound \cite{Shenker:2013pqa, Shenker:2013yza} giving support to the conjecture that they are the fastest scramblers in Nature.
The papers \cite{Murata:2017rbp, deBoer:2017xdk} investigated a Brownian particle coupled to a thermal ensemble in a holographic system. The  holographic dual object is a string hanging from the boundary of a three-dimensional BTZ black hole geometry. The other endpoint of the string is beyond the event horizon. The Lyapunov exponent can be extracted from out-of-time order four-point functions and it also saturates the chaos bound. Hence, the (sub-)system provides an example for fast scramblers which contain no gravitational degrees of freedom.

Even though the worldsheet theory contains no ordinary gravitational degrees of freedom, it shares other interesting features with theories of quantum gravity, e.g.
there are no local off-shell observables and there exist toy versions of black holes. Thus the system has been called the ``simplest theory of quantum gravity'' \cite{Dubovsky:2012wk}.

In this paper we will be concerned with the classical string.
The equation of motion of the sigma model can be re-written as a generalized sinh-Gordon equation and thus the system is integrable \cite{Pohlmeyer:1975nb,DeVega:1992xc, Bena:2003wd}
(see also the reviews \cite{Beisert:2010jr, Tseytlin:2010jv} and related papers \cite{Gubser:2002tv, Kruczenski:2004wg, Kalousios:2006xy, Alday:2007hr, Grigoriev:2007bu, Jevicki:2007aa, Jevicki:2008mm, Dorey:2008vp, Jevicki:2009uz, Dorey:2010iy, Dorey:2010id, Irrgang:2012xb}). Integrability allows for an exact discretization of the string equation of motion \cite{Vegh:2015ska, Callebaut:2015fsa, Vegh:2015yua, Gubser:2016wno, Gubser:2016zyw, Vegh:2016hwq, Vegh:2016fcm}. The corresponding embeddings are {\it segmented strings} which are exact solutions even at finite lattice-spacing. Smooth strings can be approximated by segmented strings to arbitrary accuracy (by increasing the number of segments and by choosing appropriate initial positions and velocities for the individual string segments). Segmented strings generalize piecewise linear strings in flat space \cite{Artru:1979ye, Andersson:1983ia}. The points where the segments join together move with the speed of light. This condition is necessary, otherwise the arising forces would deform the string and it would not be piecewise linear after a while.

A discrete equation of motion has been derived by the author in \cite{Vegh:2016hwq}. Precisely the same equations are satisfied when the worldsheet is coupled to a background two-form whose field strength is proportional to the volume form of AdS$_3$ \cite{Vegh:2016fcm}. (A certain value for the coupling gives the $SL(2)$ WZW model.) This non-linear discrete equation has appeared in the context of exact discretization in the mathematics literature \cite{suris}.

In this paper, we will take a smooth limit of segmented strings: we will recast the string equations of motion in terms of two fields $b(z^-, z^+)$ and $w(z^-, z^+)$ (or {\it black} and {\it white}) where $z^\pm$ are null coordinates on the worldsheet.
We will see how the $b$ and $w$ fields are related to the sinh-Gordon field. An advantage of this approach is that the string embedding can be obtained directly without solving an auxiliary scattering problem. The $b$ and $w$ fields are $\RR^{2,2}$ analogs of coordinates on the so-called {\it celestial sphere} which is the sphere at null infinity in Minkowski spacetime \cite{Pasterski:2016qvg} (see also the recent paper \cite{Atanasov:2021oyu} for the scattering problem in $(2,2)$ signature). Hence, we will sometimes refer to these fields as {\it celestial variables}.

The paper is organized as follows.
Section 2 discusses the Nambu-Goto string in AdS$_3$ spacetime and describes the discrete equation of motion that is based on boundary time fields. Section 3 introduces the new formalism by taking a continuum limit. It is shown how to compute the string embedding from the new variables and the relationship to the generalized sinh-Gordon equation is clarified. The section also discusses possible actions from which the equations of motion can be derived.
Section 4 discusses a few concrete examples where the string embeddings contain cusps. Section 5 derives the celebrated Schwarzian action using the new approach. By putting either of the $b$ or $w$ fields on-shell, an exact worldsheet action is derived which contains the Schwarzian derivative of the other field.
The paper ends with a summary of the results and an appendix in which the celestial fields are related to the spinor solutions of an auxiliary scattering problem.

\clearpage

\section{String in AdS$_3$}

Three-dimensional anti-de Sitter spacetime can be immersed into the $\RR^{2,2}$ linear ambient space. Then, AdS is the universal cover of the hyperboloid
\be
  \label{eq:hyp}
  \vec Y \cdot \vec Y \equiv -Y_{-1}^2 - Y_0^2 + Y_1^2 + Y_2^2 = -1 .
\ee
Global AdS time is the angle on the $Y_{-1}$, $Y_0$ plane. This coordinate has to be ``unwrapped'' to avoid closed time-like curves.
A part of global AdS$_3$ is covered by the \poincare patch.
The metric on this patch is
\be
  \nonumber
  ds^2 = {-dt^2 + dx^2 + dy^2 \over y^2} .
\ee
The coordinates $t, x, y$ are related to $\vec Y$ via the following transformation
\be
  \nonumber
  (t, \, x, \, y) = \le(  {Y_{0} \over Y_{2} - Y_{-1}}, \ {Y_{1} \over Y_{2} - Y_{-1}}, \ {1 \over Y_{2} - Y_{-1}} \ri),
\ee
which can be inverted on the hyperboloid:
\be
  \nonumber
  \vec Y = \le( {-1+ t^2 -x^2 - y^2 \ov 2y}, \, {t\ov y} ,  \,{x\ov y} ,  \, {1+ t^2 -x^2 - y^2  \ov 2y}\ri).
\ee
The spatial boundary of AdS lies at $y=0$. In terms of the $\vec Y$ coordinate, the boundary is the set of points that satisfy $\vec Y^2 = 0$ with the identification $\vec Y \cong c \vec Y$ (where $c \in \RR^+$).

The string can be mapped into AdS$_3$ by first taking the target space to be $\RR^{2,2}$ and then forcing the string to lie on the hyperboloid (\ref{eq:hyp}) by means of a Lagrange multiplier $\lambda$. We will work in conformal gauge. The action is given by
\be
  \label{eq:action}
  S  = -{T \ov 2}\int d\tau d\sigma ( \p_\sigma Y^\mu \p_\sigma Y_\mu - \p_\tau Y^\mu \p_\tau Y_\mu + \lambda(\vec Y^2 + 1)) ,
\ee
where $T$ is the string tension and $\vec Y(\tau, \sigma) \in \RR^{2,2}$ is the embedding function.
The non-linear equations of motion are
\be
  \label{eq:eoms}
  \p_- \p_+ \vec Y - (\p_- \vec Y \cdot \p_+ \vec Y ) \vec Y = 0 ,
\ee
where $z^- = \half(\tau-\sigma)$ and $z^+ = \half(\tau+\sigma)$ are the two null coordinates on the worldsheet and $\p_- \equiv \p_{z^-}$, $\p_+ \equiv \p_{z^+}$.
Due to the gauge choice, the equations are supplemented by the constraints
\be
  \nonumber
  \p_- \vec Y \cdot \p_- \vec Y = \p_+ \vec Y \cdot \p_+ \vec Y = 0 .
\ee

\begin{figure}[h]
\begin{center}
\includegraphics[width=5cm]{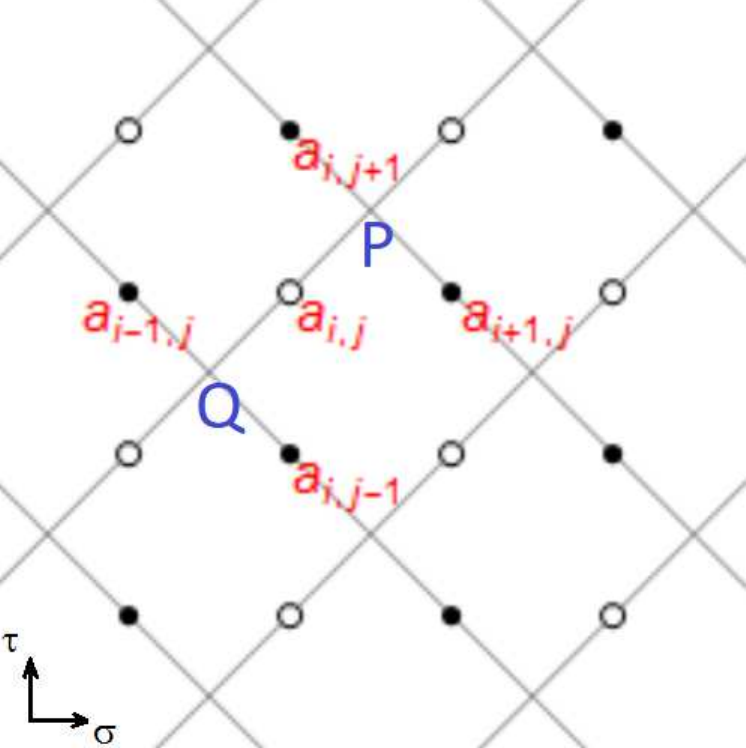}
\caption{\label{fig:sub} The string worldsheet (parametrized by $\tau$ and $\sigma$). Kink worldlines form a lattice and they collide in vertices ($P$, $Q$). The field $a_{ij}$ lives on the edges (black or white dots depending on edge orientation).
}
\end{center}
\end{figure}

\subsection{The discrete equation of motion}

The motion of the string can be equivalently described using segmented  strings \cite{Vegh:2015ska, Callebaut:2015fsa}. The classical theory is integrable which allows for the exact discretization of the equations. Solutions of the discrete equations correspond to string embeddings which consist of AdS$_2$ patches glued along null rays: the worldlines of kinks.
Kink worldlines form a quad lattice on the worldsheet, see FIG. \ref{fig:sub}. Each diamond in the figure is a patch of AdS$_2$ with a constant normal vector which is defined as
\be
  \nonumber
  N_a(z^-, z^+) \propto \epsilon_{abcd} Y^b \p_- Y^c \p_+ Y^d \in \RR^{2,2} \qquad \vec N^2 = 1
\ee
Indices are lowered and raised  by $\eta = \textrm{diag}(-1,-1,+1,+1)$.
The normal vector satisfies the same equation of motion as $\vec Y$ in (\ref{eq:eoms}), but with a plus sign \cite{Vegh:2015ska, Callebaut:2015fsa, Vegh:2016hwq, Gubser:2016wno}.

A discrete evolution equation for the normal vectors (or, equivalently, the kink collision points) has been found \cite{Vegh:2015ska, Callebaut:2015fsa} and can be used to build segmented string solutions.
In \cite{Vegh:2016hwq} I showed that segmented strings in AdS$_3$ move according to the discrete equation of motion
\be
  \label{eq:deqn}
  \hskip -0.15cm {1\ov a_{ij} - a_{i,j+1}}+   {1\ov a_{ij} - a_{i,j-1}} =
  {1\ov a_{ij} - a_{i+1,j}}+   {1\ov a_{ij} - a_{i-1,j}} .
\ee
Here $i$ and $j$ are integer indices labeling lattice points on the string worldsheet. As illustrated in FIG. \ref{fig:sub}, kink worldlines pass through each of the lattice points (black and white dots). Kinks move with the speed of light both in target space and on the worldsheet.
The dots are colored alternatingly, depending on which way the kink moves.

\clearpage

In FIG. \ref{fig:sub}, the variable $a_{ij}$ is expressed using the components of the difference vector $V \equiv P-Q  \in \RR^{2,2} $. If we define the $ \mathfrak{a}: \RR^{2,2} \to \RR$ function as
\be
  \mathfrak{a}(\vec X) \equiv {X_{-1} + X_2 \over X_0 + X_1}
\ee
then we simply have
\be
a_{ij} = \mathfrak{a}(\vec V)
\ee
and similarly the other $a$ variables are computed from their respective difference vectors. We will see shortly that they correspond to (advanced or retarded) boundary times.

Note that the equation of motion (\ref{eq:deqn}) is invariant under Möbius transformations,
\be
  \nonumber
  a \to {c_1 + c_2 a \ov c_3 + c_4 a}
\ee
which is due to the left $SL(2)$ factor in the AdS$_3$ isometry group $SO(2,2) = SL(2) \times SL(2)$.
The $a$ field does not completely specify the string embedding: the right $SL(2)$ group only acts on the ``right-handed'' variables, which will be denoted by a tilde:
\be
\nonumber
\tilde a_{ij} = \mathfrak{\tilde a}(\vec V)  \equiv {V_{-1} + V_2 \over -V_0 + V_1} \, .
\ee
The $\tilde a$ field satisfies the same  equation as $a$ in (\ref{eq:deqn}).

What is the meaning of the $a$ field? The kink difference vectors (i.e. $V$ above) are null, therefore they correspond to points where the rays hit the boundary of AdS$_3$.
In terms of the boundary \poincare coordinates $t$ and $x$, the difference vector can be expressed as
\be
  \nonumber
  \vec V \propto   \left( \begin{array}{c}
    {-1+ t^2 -x^2  }  \\
     2t  \\
     2x  \\
     {1+ t^2 -x^2 }
   \end{array} \right).
\ee
from which we get
\bea
  \nonumber
  a &=& x^-   \\
  \nonumber
  \tilde a &=& -x^+
\eea
where $x^\pm \equiv t \pm x$.
If we now consider an embedding of AdS$_2 \subset$ AdS$_3$ with $x=0$ on the \poincare patch, we see that the $a$ field tells us the retarded and advanced times at which kink null rays hit the boundary.

\clearpage

\begin{figure}[h]
\begin{center}
\includegraphics[scale=0.33]{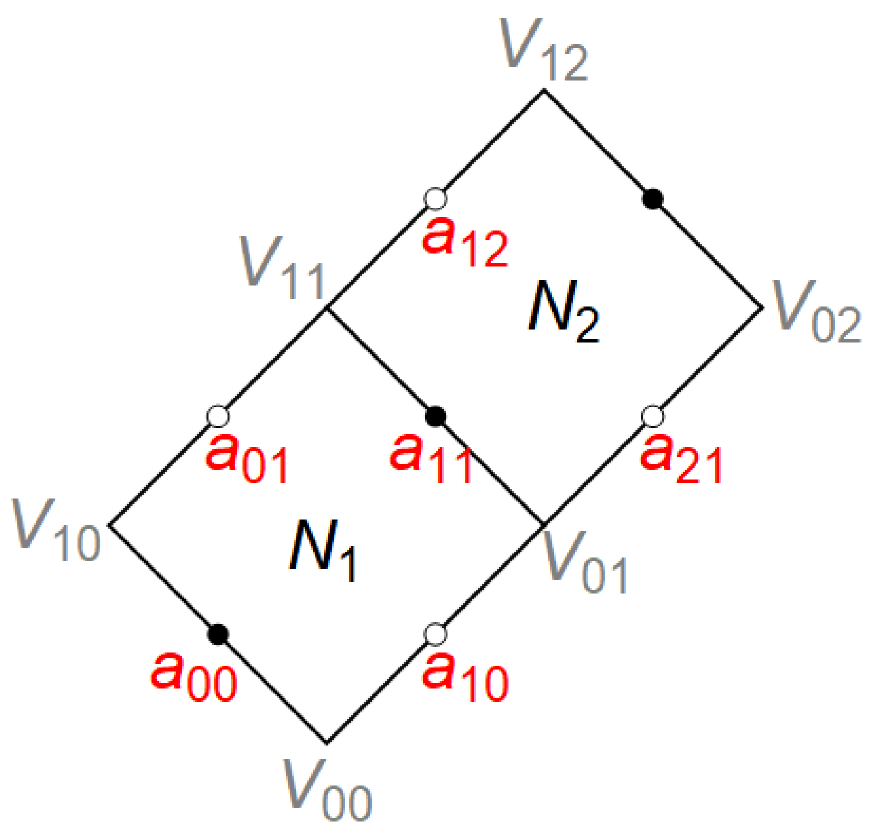} \caption{\label{fig:double}
Two adjacent patches on the worldsheet.}
\end{center}
\end{figure}

\section{Celestial fields}

In this section we derive partial differential equations by taking an appropriate continuum limit of the discrete equation of motion (\ref{eq:deqn}).

\subsection{Continuum limit}

In general there are different ways to take a continuum limit of discrete equations. In the following we investigate the case where the $a$ variables over black dots and white dots (see FIG. \ref{fig:sub}) converge to {\it two} distinct fields. We will denote them by $b(z^-, z^+)$ and  $w(z^-, z^+)$ and call them {\it black} and {\it white} celestial fields, respectively\footnote{They could be called retarded and advanced fields in principle, but these labels depend on which direction one follows the null rays (i.e. whether one considers the left or the right boundary of the AdS$_2$ patch).}.

Let us consider two adjacent patches as in FIG. \ref{fig:double} and set
\bea
  \nonumber
  a_{11} &=& b(z^-, z^+)  \\
  \nonumber
  a_{00} &=& b(z^-,   z^+ -2\epsilon) \\
  \nonumber
  a_{10} &=& w(z^- -\epsilon,   z^+ -\epsilon) \\
  \nonumber
  a_{12} &=& w(z^- +\epsilon,   z^+ +\epsilon) \\
  \nonumber
  a_{01} &=& w(z^- +\epsilon,   z^+ -\epsilon) \\
  \label{eq:adef}
  a_{21} &=& w(z^- -\epsilon,   z^+ +\epsilon)
\eea
By taking the $\epsilon \to 0$ small lattice spacing  limit, the discrete equation (\ref{eq:deqn}) becomes
\be
  \label{eq:eomwhite}
  \p_- \p_+ w = {2 \p_- w \, \p_+ w \over w-b }
\ee
and similarly for the black field
\be
  \label{eq:eomblack}
  \p_- \p_+ b = {2 \p_- b \, \p_+ b \over b-w } .
\ee

In the continuum limit, the two fields can be computed via
\be
  \nonumber
  b = \mathfrak{a}(\p_- \vec Y) \qquad \textrm{and} \qquad w = \mathfrak{a}(\p_+ \vec Y)
\ee
where $\vec Y$ is the embedding function from (\ref{eq:action}).
Note that the black and white fields transform as scalars under worldsheet conformal transformations.

\subsection{Area densities $e^\alpha$ and $e^\beta$}

Let us define the worldsheet area density $e^\alpha$ and the dual area density $e^\beta$ as follows:
\be
  \nonumber
  \alpha(z^-, z^+) = \log \p_- \vec Y \cdot \p_+ \vec Y
\ee
\be
   \nonumber
 \beta(z^-, z^+) = \log \p_- \vec N \cdot  \p_+ \vec N
\ee
We can determine the area of an elementary patch of a segmented string \cite{Vegh:2016hwq}. For instance, the area of patch \#1 in FIG. \ref{fig:sub} is given by
\be
  \nonumber
  A= \log\le( { (a_{00}-a_{10})(a_{01}-a_{11}) \ov (a_{00}-a_{01})(a_{10}-a_{11})}  \ri)^2 = e^\alpha (2\epsilon)^2
\ee
If we set the $a$ variables according to (\ref{eq:adef}), then in the $\epsilon \to 0 $ limit we get
\be
  \label{eq:alpha}
  e^{\alpha} = 2{\p_+ b \, \p_- w  \over (b-w)^2} \qquad
  e^{\beta} =  2{ \p_- b \, \p_+ w \over (b-w)^2} .
\ee
$\alpha$ is an important field, since as we will see it satisfies the generalized sinh-Gordon equation, and its exponential is the Nambu-Goto Lagrangian. Note that $e^{\alpha}$ is always non-negative whereas $e^{\beta}$ can be negative in certain regions on the worldsheet.

\subsection{Target space vs. dual space}

Let us consider two adjacent AdS$_2$ patches as in FIG. \ref{fig:double}. If the normal vector $\vec N_1$ and the $a$ variables are known, then the vertices can be computed as follows. Let us define the antisymmetric matrix
\be
\nonumber
M^{\mu\nu}_{(a,b)} = {1\ov a-b}\left(
\begin{array}{cccc}
 0 & -1-a b & 1-a b & a+b \\
 1+a b & 0 & -a-b & 1-a b \\
 a b-1 & a+b & 0 & -1-a b \\
 -a-b & a b-1 & 1+a b & 0 \\
\end{array}
\right)
\ee
where $a$ and $b$ are two parameters.
Note that
\be
  M^{\mu\nu}_{(a,b)} = - M^{\mu\nu}_{(b,a)}  \qquad \textrm{and} \qquad  {M^\mu}_\nu {M^\nu}_\kappa = \delta^\mu_\kappa
\ee
where indices are raised by $\eta = \textrm{diag}(-1,-1,+1,+1)$.

Then the vertices are given by
\bea
  \nonumber
  (V_{00})^\mu &=& M^{\mu\nu}_{(a_{00},a_{10})}  (N_1)_\nu   \\
  \label{eq:vertices}
  (V_{01})^\mu &=& M^{\mu\nu}_{(a_{11},a_{10})}  (N_1)_\nu
\eea
and similarly for the other vertices: the two parameters of the $M$ matrix are the $a$ values sitting on the black and white dots near the vertex (see FIG. \ref{fig:sub}).
Since $  M^{\mu\nu}M_{\mu\kappa} =  -\delta^\nu_\kappa $ and $N_1^2 = 1$, the vertices will be on the AdS hyperboloid, i.e. $V_{00}^2 = V_{01}^2 = -1$. Furthermore, since $M$ is antisymmetric, we will also have $V_{00} \cdot N_1 = V_{01} \cdot N_1 = 0$, which means that they lie on the AdS$_2$ patch defined by $N_1$. Finally, it is easily shown that $ \mathfrak{a}(V_{01}-V_{00}) = a_{10}$. This means that the expressions for the vertices in (\ref{eq:vertices}) are indeed correct.

It can further be shown that
\be
  \nonumber
  (N_2)^\mu = M^{\mu\nu}_{(a_{11},a_{21})}  (V_{01})_\nu \ .
\ee
Thus, if the $a$ variables are known, then the $M$ matrix can be used to switch between target space and dual space.

One can define an analogous antisymmetric matrix which does the same ``reflection'' operation, but this time using the tilde variables:
\be
\nonumber
\widetilde{M}^{\mu\nu}_{(\tilde a,\tilde b)} = {1\ov \tilde a-\tilde b}\left(
\begin{array}{cccc}
 0 & -1-\tilde a \tilde b & -1+\tilde a \tilde b & -\tilde a-\tilde b \\
 1+\tilde a \tilde b & 0 & -\tilde a-\tilde b & 1-\tilde a \tilde b \\
 1-\tilde a \tilde b & \tilde a+\tilde b & 0 & 1+\tilde a \tilde b \\
 \tilde a+\tilde b & \tilde a \tilde b-1 & -1-\tilde a \tilde b & 0 \\
\end{array}
\right) .
\ee
Then we have
\be
  \label{eq:vertices2}
  (V_{00})^\mu = \widetilde{M}^{\mu\nu}_{(\tilde a_{00},\tilde a_{10})}  (N_1)_\nu \ .
\ee

In the continuum limit, the $a$ variables converge to the $b$ and $w$ fields. Hence,
\be
  \nonumber
   Y^\mu(z^-, z^+) =M^{\mu\nu}_{(b,w)} N_\nu(z^-, z^+) \ ,
\ee
\be
  \nonumber
   N^\mu(z^-, z^+) =M^{\mu\nu}_{(b,w)} Y_\nu(z^-, z^+) \ .
\ee

\subsection{The auxiliary $u$ and $v$ fields}

Let us define the following quantities
\be
  \nonumber
  u \equiv \p_- N \cdot\p_- Y = -N\cdot \p_-\p_- Y  =  - Y \cdot \p_-\p_- N
\ee
\be
  \nonumber
  v \equiv  -\p_+ N \cdot\p_+ Y = N\cdot \p_+\p_+ Y  =  Y \cdot \p_+\p_+ N
\ee
We need to find a properly discretized versions of these equations.
If we set the $a_{ij}$ variables as in (\ref{eq:adef}), and take
\be
  \nonumber
  v \approx -(N_2 - N_1) \cdot (V_{01}- V_{00}) {1\ov (2\epsilon)^2 }
\ee
then in the $\epsilon \to 0 $ limit we get
\be
  \nonumber
  v( z^+) =  2{  \p_+ b  \, \p_+ w \over (b-w)^2}
\ee
A similar calculation gives
\be
  \nonumber
  u(z^-) =  2{ \p_- b \, \p_- w \over (b-w)^2} \qquad
\ee
It is known that $ \p_+ u = \p_- v = 0$. In fact, these two equations are equivalent to  (\ref{eq:eomwhite}) and (\ref{eq:eomblack}).

\subsection{The generalized sinh-Gordon equation}

By plugging in the expressions for $\alpha$, $u$, and $v$, it is easy to see that $\alpha$ satisfies the generalized sinh-Gordon equation
\be
  \label{eq:sinhg}
   \p_- \p_+ \alpha + e^{\alpha} - uv  e^{-\alpha} = 0
\ee
This equation was first derived in \cite{Pohlmeyer:1975nb} (see also \cite{DeVega:1992xc}).
Using
\be
   \nonumber
  e^{\alpha + \beta} = uv
\ee
the sinh-Gordon equation can be re-written as
\be
   \nonumber
  \p_- \p_+ \alpha + e^{\alpha} - e^{\beta} = 0
\ee
The dual field $\beta$ satisfies the same equation with $\alpha$ and $\beta$ exchanged:
\be
   \nonumber
  \p_- \p_+ \beta + e^{\beta} - e^{\alpha} = 0
\ee
In a worldsheet region where $uv>0$ one can perform a conformal transformation and (locally) set $u(z^-) = v(z^+)  = 1$. We will call these {\it balanced coordinates} on the worldsheet. In these  coordinates we have $\alpha = -\beta$. Note that these coordinates typically do not cover the entire worldsheet since $uv$ can change sign.

Finally, note that $\alpha$ can be expressed from $b$ and $u$ or $w$ and $v$:
\be
  \label{eq:expressa}
  e^{\alpha} =  {u  }{ \p_+ b  \over  \p_- b} =  {v  }{ \p_- w  \over  \p_+ w}
\ee
These formulas simplify even further in balanced coordinates when $u=v=1$.

\subsection{Constraints}
\label{sec:con}

The discussion has so far focused on the continuum limit of the $a_{ij}$ field while neglecting the $\tilde a_{ij}$ variables and the corresponding $\tilde b$ and $\tilde w$ fields.

The $a$ and $\tilde a$ fields are not independent which can be seen as follows. Let us consider a single AdS$_2$ patch (see patch \#1 in FIG \ref{fig:double}). The patch is bounded by four kink worldlines.
These are null vectors in $\RR^{2,2}$ which can be constructed from the $a$ and $\tilde a$ variables. For instance,
\bea
   \nonumber
   \vec p^{(1)} &\equiv& \le( -1-a_{00} \tilde a_{00}, \, a_{00}-\tilde a_{00} ,  \, -a_{00}-\tilde a_{00} ,  \, 1- a_{00} \tilde a_{00}\ri) \ \propto \ \vec V_{10}- \vec V_{00}  \\
   \nonumber
   \vec p^{(2)} &\equiv& \le( -1-a_{10} \tilde a_{10}, \, a_{10}-\tilde a_{10} ,  \, -a_{10}-\tilde a_{10} ,  \, 1- a_{10} \tilde a_{10}\ri)  \ \propto \  \vec V_{01}- \vec V_{00}  \\
   \nonumber
   \vec p^{(3)} &\equiv& \le( -1-a_{11} \tilde a_{11}, \, a_{11}-\tilde a_{11} ,  \, -a_{11}-\tilde a_{11} ,  \, 1- a_{11} \tilde a_{11}\ri)  \ \propto \  \vec V_{11}- \vec V_{01}  \\
  \label{eq:vdef}
   \vec p^{(4)} &\equiv& \le( -1-a_{01} \tilde a_{01}, \, a_{01}-\tilde a_{01} ,  \, -a_{01}-\tilde a_{01} ,  \, 1- a_{01} \tilde a_{01}\ri)  \ \propto \  \vec V_{11}- \vec V_{10}
\eea
It is easy to check that e.g. $ \mathfrak{a}(\vec p^{(1)}) = a_{00}$ and $ \mathfrak{\tilde a}(\vec p^{(1)}) = \tilde a_{00}$ and similarly for the other vectors.
Since the sum of the four difference vectors (with appropriate signs) trivially vanishes, the determinant of the $4\times 4$ matrix constructed from the $p^{(i)}$ as row vectors must also vanish
\be
  \label{eq:condef}
  \det
  \left( \begin{array}{cccc}
     p^{(1)}_{-1} & p^{(1)}_{0} &  p^{(1)}_{1} &  p^{(1)}_{2}  \\
     p^{(2)}_{-1} & p^{(2)}_{0} &  p^{(2)}_{1} &  p^{(2)}_{2}  \\
     p^{(3)}_{-1} & p^{(3)}_{0} &  p^{(3)}_{1} &  p^{(3)}_{2}  \\
     p^{(4)}_{-1} & p^{(4)}_{0} &  p^{(4)}_{1} &  p^{(4)}_{2}
   \end{array} \right) = 0
\ee
This is one scalar constraint equation for the eight $a$ and $\tilde a$ variables around a patch. The remaining seven degrees of freedom completely specify the patch. This can be seen as follows. Let us fix the values of the $a$ and $\tilde a$ variables and keep $N_1$ arbitrary for the moment. Then four vertices $V_{00}$, $V_{10}$, $V_{01}$, and $V_{11}$ can be determined from (\ref{eq:vertices}) and four other vertices $\tilde V_{00}$, $\tilde V_{10}$, $\tilde V_{01}$, and $\tilde V_{11}$ using the analogous equation (\ref{eq:vertices2}).
Now if we set $V_{00}=\tilde V_{00}$ and $V_{10}=\tilde V_{10}$, then these equations completely determine the components of $N_1$. We get
\be
 \label{eq:forn}
 \hskip -0.3cm \vec N_1= {\tiny {1\ov \mathcal{C}} \left( \begin{array}{c}
  -a_{01} \tilde a_{10} (a_{00} (\tilde a_{01}-\tilde a_{00})+a_{10} \tilde a_{00}+1)+a_{10} (-\tilde a_{00} (a_{00} \tilde a_{01}+1)+a_{00} \tilde a_{01} \tilde a_{10}+\tilde a_{01})+a_{00} (\tilde a_{10}-\tilde a_{01})+a_{01} \tilde a_{00} (a_{10} \tilde a_{01}+1) \\
 a_{00} (a_{01} (\tilde a_{01}-\tilde a_{00})+a_{10} \tilde a_{00}-a_{10} \tilde a_{10}-\tilde a_{00} \tilde a_{01}+\tilde a_{00} \tilde a_{10})+a_{01} (-a_{10} \tilde a_{01}+a_{10} \tilde a_{10}+\tilde a_{00} \tilde a_{01}-\tilde a_{01} \tilde a_{10})+a_{10} \tilde a_{10} (\tilde a_{01}-\tilde a_{00}) \\
  a_{00} (a_{01} (\tilde a_{00}-\tilde a_{01})-\tilde a_{00} (a_{10}+\tilde a_{01})+\tilde a_{10} (a_{10}+\tilde a_{00}))+a_{01} \tilde a_{01} (a_{10}+\tilde a_{00})-a_{01} \tilde a_{10} (a_{10}+\tilde a_{01})+a_{10} \tilde a_{10} (\tilde a_{01}-\tilde a_{00}) \\
  a_{01} (\tilde a_{00} (a_{00} \tilde a_{10}+a_{10} \tilde a_{01}-a_{10} \tilde a_{10}-1)-a_{00} \tilde a_{01} \tilde a_{10}+\tilde a_{10})+a_{10} (-a_{00} \tilde a_{00} \tilde a_{01}+a_{00} \tilde a_{01} \tilde a_{10}+\tilde a_{00}-\tilde a_{01})+a_{00} (\tilde a_{01}-\tilde a_{10})
   \end{array} \right)
}
\ee
where the normalization factor is
\be
  \nonumber
 \mathcal{C}^2 \equiv 4(a_{01}-a_{00}) (a_{00}-a_{10}) (a_{01}-a_{10}) (\tilde a_{00}-\tilde a_{01}) (\tilde a_{00}-\tilde a_{10}) (\tilde a_{01}-\tilde a_{10}) .
\ee

Finally, for the other two vertices we get $V_{01}=\tilde V_{01}$ and $V_{11}=\tilde V_{11}$ if and only if the determinant (\ref{eq:condef}) vanishes.

There is an analogous dual constraint for the difference vectors computed from four adjacent normal vectors.
It is easy to see that in the continuum limit (\ref{eq:condef}) and the dual constraint are tantamount to
\be
  \nonumber
  \alpha =  {\tilde \alpha } \qquad \textrm{and} \qquad   \beta =  {\tilde \beta }
\ee
where the $ \tilde \alpha$ and $\tilde \beta$ are computed from the $\tilde b$ and $\tilde w$ fields:
\be
  \nonumber
  e^{\tilde \alpha} = 2{\p_+ \tilde b \, \p_- \tilde w  \over (\tilde b-\tilde w)^2} \qquad
  e^{\tilde \beta} =  2{ \p_- \tilde b \, \p_+ \tilde w \over (\tilde b-\tilde w)^2}
\ee

Although it is not a separate constraint, one can derive an equation for $\tilde u$ and $\tilde v$ as well. The string equation of motion (\ref{eq:eoms}) says that $\vec Y \propto \p_- \p_+ Y$. The normal vector satisfies an analogous equation of motion (with a plus sign). Thus we have
\be
  \label{eq:merol}
  0 = \vec Y \cdot \vec N  \propto (\p_- \p_+ Y) \cdot (\p_- \p_+ N)
\ee
In order to express the right-hand side, similarly to (\ref{eq:vdef}), let us now take the following ans\"atze
\be
 \label{eq:embeq}
   \p_+ \vec Y= \lambda(z^-,z^+)
  \left( \begin{array}{c}
    -1-w \tilde w  \\
     w-\tilde w  \\
     -w-\tilde w  \\
     1- w \tilde w
   \end{array} \right), \qquad
 \p_- \vec N= \kappa(z^-,z^+)
  \left( \begin{array}{c}
    -1-w \tilde w  \\
     w-\tilde w  \\
     -w-\tilde w  \\
     1- w \tilde w
   \end{array} \right)
\ee
with arbitrary proportionality factors $\lambda(z^-,z^+)$ and $\kappa(z^-,z^+)$. If we plug these expressions into (\ref{eq:merol}), we get
\be
  \nonumber
   2 \lambda\kappa( \p_- w \, \p_+ \tilde w + \p_+ w \, \p_- \tilde w) =0 .
\ee
An analogous equation containing $b$ and $\tilde b$ can be derived if we start instead with a similar ansatz for $\p_- \vec Y$ and $ \p_+ \vec N$.
Combining these equations with (\ref{eq:expressa}) we get the constraints
\be
  \nonumber
  \tilde u = -u \qquad \tilde v = -v .
\ee

In addition to the above constraints, the area density must also be non-negative. This is ensured if the black and white fields satisfy
\be
  \label{eq:dbdwcon}
  \p_+ b \, \p_- w \ge 0 .
\ee

\subsection{String embedding}

Given a solution of the sinh-Gordon equation, the string embedding can be computed by solving an auxiliary Dirac equation where $\alpha$ appears as a potential. If the $b$, $\tilde b$, $w$, $\tilde w$ fields are known, there is a simpler direct way to compute the embedding.

Similarly to (\ref{eq:vdef}), we can take the following ansatz for $\p_+ \vec Y$:
\be
 \label{eq:embeq}
  \p_+ \vec Y= \lambda(z^-, z^+)
  \left( \begin{array}{c}
    -1-w \tilde w  \\
     w-\tilde w  \\
     -w-\tilde w  \\
     1- w \tilde w
   \end{array} \right).
\ee
The proportionality factor $\lambda(z^-, z^+)$ can be determined as follows. From the string equation of motion (\ref{eq:eoms}) we have
\be
  \nonumber
    \vec Y = e^{-\alpha} \p_- \p_+ \vec Y
\ee
and $\alpha$ can be expressed using $b$ and $w$ as in (\ref{eq:alpha}).
If we plug in the $z$-derivative of (\ref{eq:embeq}), we can compute the norm of the position vector. We get
\be
  \nonumber
  \vec Y^2 = 4 e^{-2\alpha} \lambda^2  \p_- w  \, \p_- \tilde w
\ee
Note that terms containing partial derivatives of $\lambda$ have dropped out.
Since the string must lie on the AdS$_3$ hyperboloid, we have $\vec Y^2 = -1$, which determines $\lambda$ and therefore also the string embedding.
However, the resulting expression is complicated and contains second derivatives of $b$ and $w$. A simpler formula can be obtained by taking the continuum limit of (\ref{eq:forn}) and then converting it into a target space vector via (\ref{eq:vertices}). We get
\be
  \nonumber
 \vec Y= \pm {1\ov \mathcal{D}}
  \left( \begin{array}{c}
    (\tilde w - \tilde b)(1+ b \tilde w) \p_- w + (w-b)(1+\tilde b w) \p_- \tilde w  \\
    (b - \tilde w)(\tilde b - \tilde w) \p_- w + (b-w)(w- \tilde b) \p_- \tilde w   \\
    ( \tilde w - \tilde b)(b + \tilde w) \p_- w + (w-b)( \tilde b+w) \p_- \tilde w   \\
    (\tilde b - \tilde w)(1- b \tilde w) \p_- w + (b-w)(1-\tilde b w) \p_- \tilde w
   \end{array} \right)
\ee
where the normalization factor is defined by
\be
  \nonumber
 \mathcal{D}^2 \equiv { -4(b-w)^2 (\tilde b - \tilde w)^2 \p_- w \, \p_- \tilde w } .
\ee

\subsection{The action}

The Nambu-Goto action (after setting the prefactor to one)
\be
  \nonumber
  S = \int\sqrt{-g} \, dz^- dz^+ =   \int e^{\alpha}  dz^- dz^+
\ee
If we now plug in the expression (\ref{eq:alpha}) for $e^{\alpha}$, then we obtain the following action
\be
  \nonumber
  S =   2 \int {\p_+ b \, \p_- w  \over (b-w)^2}   dz^- dz^+
\ee
The Euler-Lagrange equations for $b$ and $w$ give the equations of motion (\ref{eq:eomwhite}) and (\ref{eq:eomblack}). Note that only those solutions are allowed that satisfy (\ref{eq:dbdwcon}).

The same equations of motion are obtained from the ``dual'' action
\be
  \nonumber
  S' =     \int e^{\beta}  dz^- dz^+ = 2 \int {\p_+ w \, \p_- b  \over (b-w)^2}  dz^- dz^+ \, .
\ee
$S'$ also appears in expressions for the regularized area of the worldsheet which equals the expectation value of the dual Wilson loop at strong coupling (see e.g. Appendix B of \cite{Kruczenski:2014bla}).

\subsection{Various limits}

\subsubsection{Flat space limit}
\label{sec:flat}

In the flat space limit, the string worldsheet is mapped into an infinitesimal volume of AdS$_3$. Thus, the area density $e^{\alpha}$ vanishes and the generalized sinh-Gordon equation (\ref{eq:sinhg}) degenerates into the Liouville equation
\be
  \nonumber
   \p_- \p_+ \alpha  - uv  e^{-\alpha} = 0
\ee
This equation can be  explicitly solved in terms of two functions $f(z^-)$ and $g(z^+)$
\be
  \label{eq:liousol}
   e^{\alpha} = { u v \ov 2} {(f-g)^2 \ov f'g'}
\ee

This solution can be obtained using the black and white fields if we consider the limit
\be
  \nonumber
  b(z^-,  z^+) = b_0(z^-) + \epsilon b_1(z^-,  z^+) + \mathcal{O}(\epsilon^2) \qquad
  w(z^-,  z^+) = w_0(z^+) + \epsilon w_1(z^-,  z^+)   + \mathcal{O}(\epsilon^2)
\ee
as $\epsilon \to 0$. Here $b_0$ and $w_0$ are arbitrary functions, whereas $b_1$ and $w_1$ are determined from the equations of motion. Then, $\alpha$ can be determined from (\ref{eq:alpha})
\be
   \nonumber
 e^\alpha = uv e^{-\beta} = {u v \ov 2} {(b_0 - w_0)^2 \ov b_0' w_0' }  + \mathcal{O}(\epsilon^2)
\ee
Comparing this with (\ref{eq:liousol}) gives $f= b_0$ and $g=w_0$.

\subsubsection{AdS$_2$ limit}

In the flat space limit one had $X(z^-,  z^+) \approx X_0$ for a fixed $X_0 \in \RR^{2,2}$ vector in target space. One can consider a similar limit in the dual space of normal vectors, i.e. $N(z^-,  z^+) \approx N_0$.
In this case, the black and white fields become
\be
  \nonumber
  b(z^-,  z^+) \approx b_0(z^+)  \qquad \textrm{ and} \qquad w(z^-,  z^+) \approx w_0(z^-)
\ee
Note that the dependence of $b$ and $w$ on the worldsheet coordinates is swapped compared to the flat space limit.

\subsubsection{Non-linear waves moving in one direction}

One can consider non-linear waves moving in one direction on the string in AdS$_3$. In \cite{Mikhailov:2003er} Mikhailov gave an explicit expression for such solutions in terms of the position function of the string endpoint on the boundary. On the \poincare patch the solution is
\bea
  \nonumber
  t(\tau, y) &=& \tau + {y\ov \sqrt{1-x_0'(\tau)^2}} \\
  \nonumber
  x(\tau, y) &=& x_0(\tau) + {y x_0'(\tau)\ov \sqrt{1-x_0'(\tau)^2}} \\
  y(\tau, y) &=& y
\eea
where $x_0(\tau)$ specifies the endpoint of the string in terms of the retarded time $\tau$.
The induced metric is locally AdS$_2$ everywhere. The corresponding black and white fields can be computed with the result
\be
  \nonumber
  b  = b(z^-,  z^+) \qquad
  w  = w_0(z^-) 
\ee
We see that only $w$ has a special form. These solutions describe right-moving waves. For left-moving waves one has
\be
  \nonumber
  b  = b_0(z^+) \qquad
  w  = w(z^-,  z^+) 
\ee

Finally, one can consider a ``dual'' non-linear wave limit in which
\be
  \nonumber
  b  = b(z^-,  z^+) \qquad
  w  = w_0(z^+) 
\ee
and $b  = b_0(z^-)$, $w  = w(z^-,  z^+)$ for its left-moving counterpart.


\section{Examples}

The generalized sinh-Gordon equation in balanced coordinates gives the ordinary sinh-Gordon equation. This equation has singular soliton and antisoliton solutions, e.g.
\be
  \label{eq:solitt}
  {\alpha}_{s,\bar{s}}= \pm \log \tanh^2  {z^- - z^+ \ov \sqrt{2}}
\ee
There are explicit formulas for solutions containing solitons. The corresponding string embeddings have also been calculated, see \cite{Kruczenski:2004wg, Jevicki:2007aa, Jevicki:2008mm, Jevicki:2009uz}. The string has a cusp whenever the  $e^\alpha$ area density vanishes. This happens precisely at the location of a soliton.

In this section, we compute the black and white fields for a few examples.

\subsection{Rotating string}

The embedding is given by \cite{Gubser:2002tv, Jevicki:2007aa}
\be
  \nonumber
 \vec Y(\tau, \sigma)=
  \left( \begin{array}{c}
     \cosh \sigma \sin \tau  \\
    - \cosh \sigma \cos \tau  \\
    \sinh \sigma \cos \tau  \\
    \sinh \sigma \sin \tau
   \end{array} \right).
\ee
From this, we compute
\be
  \nonumber
  b(z^-,  z^+) = e^{z^- - z^+} \qquad
  w(z^-,  z^+) = -e^{z^- -  z^+}
\ee
\be
  \nonumber
  \tilde b(z^-,  z^+) = {2\ov 1- \tan {z^- + z^+ \ov 2}} -1 \qquad
  \tilde w(z^-,  z^+) = 1-{2\ov 1+ \tan {z^- +  z^+ \ov 2}}
\ee
and the auxiliary fields:
\be
  \nonumber
  e^\alpha = e^{\tilde \alpha} = e^\beta = e^{\tilde \beta} = -u = \tilde u = -v = \tilde v = \half .
\ee

\clearpage

\begin{figure}[h]
\begin{center}
\includegraphics[scale=0.6]{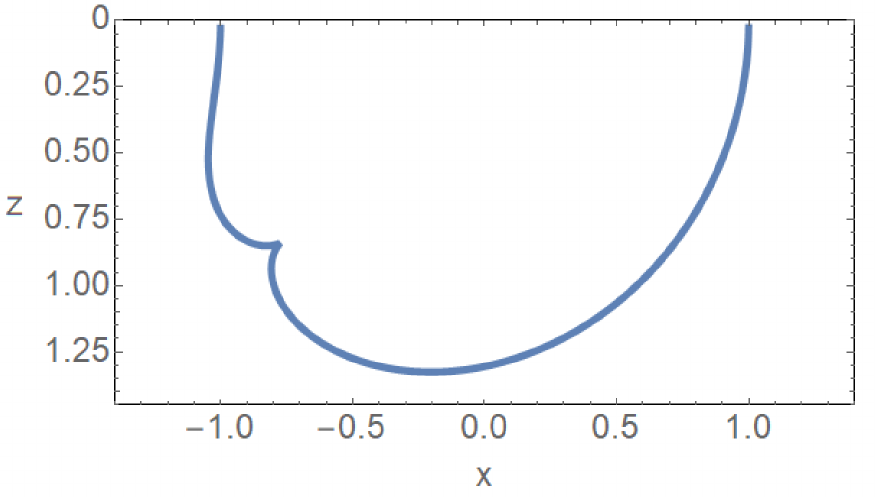} \caption{
An example solution: a time-slice of the one-cusp solution on the \poincare patch. The endpoints and the cusp move with the speed of light on the boundary and in the bulk, respectively.}
\end{center}
\end{figure}

\subsection{String with one cusp}

The embedding is given by \cite{Jevicki:2007aa}
\be
  \nonumber
 \vec Y(\tau, \sigma)={1\ov 2 \sqrt{2}\cosh \sigma}
  \left( \begin{array}{c}
    2\tau \cos\tau - (2+\cosh 2\sigma)\sin \tau  \\
    2\tau \sin\tau -(2+\cosh 2\sigma) \cos \tau  \\
    -2\tau \cos \tau + \sin \tau \cosh 2\sigma  \\
     -2\tau \sin \tau - \cos \tau \cosh 2\sigma
   \end{array} \right).
\ee

The black and white fields are
\be
  \nonumber
  b(z^-,  z^+) = {1+z^- + z^+ - \sinh(z^- - z^+) \over 1-z^- - z^+ + \sinh(z^- - z^+)}
\qquad   w(z^-,  z^+) = {1+z^-  + z^+ + \sinh(z^- - z^+) \over 1-z^- - z^+ - \sinh(z^- -  z^+)}
\ee
\be
  \nonumber
  \tilde b(z^-,  z^+) = {e^{z^-} + e^{z^+} \tan {z^- +  z^+ \ov 2} \over e^{z^+} - e^{z^-} \tan {z^- + z^+ \ov 2}}
\qquad
  \tilde w(z^-,  z^+) = { e^{ z^+} + e^{z^-} \tan {z^- +  z^+ \ov 2} \over
e^{z^-} - e^{z^+} \tan {z^- + z^+ \ov 2} }
\ee
We further have
\be
  \nonumber
  -u = \tilde u = -v = \tilde v = \half  ,
\ee
and
\be
  \nonumber
  e^\alpha = e^{\tilde \alpha} = {1\ov 4e^{\beta}} = {1\ov 4e^{\tilde \beta}} =  \half \tanh^2  {z^- -  z^+ \ov 2}
\ee
which, in balanced coordinates, is nothing but the soliton solution (\ref{eq:solitt}).

\subsection{String with two cusps}

The embedding with two cusps is given via the following two complex functions \cite{Jevicki:2007aa}
\be
  \nonumber
  Y_{-1}+ i Y_0 = e^{-i \tau}{  v \cosh T \cosh \sigma + \cosh X \cosh \sigma - \gamma^{-1} \sinh X \sinh \sigma + i \gamma^{-1} \sinh T \cosh \sigma   \ov \cosh T + v \cosh X}
\ee
\be
  \nonumber
  Y_{1}+ i Y_2 = e^{-i \tau}   { v \cosh T \sinh \sigma + \cosh X \sinh \sigma - \gamma^{-1} \sinh X \cosh \sigma + i \gamma^{-1} \sinh T \sinh \sigma  \ov \cosh T + v \cosh X}
\ee
where $\gamma^{-1} = \sqrt{1-v^2}$ is the Lorentz factor, $T = 2v \gamma \tau$, and $X = 2 \gamma \sigma$. Finally $v$ is the asymptotic speed of the two cusps on the worldsheet in the center-of-mass frame.

For the black and white fields we get

\be
  \nonumber
  b(z^-,  z^+) = {\gamma^{-1} \sinh {z^- -  z^+ \ov 2} \cosh \gamma(z^- -  z^+) -  \cosh{z^- -  z^+ \ov 2} \le[ \sinh\gamma(z^- -  z^+) + \sinh v\gamma(z^- +  z^+) \ri]  \ov
\gamma^{-1} \cosh {z^- -  z^+ \ov 2} \cosh \gamma(z^- -  z^+) -  \sinh{z^- -  z^+ \ov 2} \le[ \sinh\gamma(z^- -  z^+) + \sinh v\gamma(z^- +  z^+) \ri]}
\ee
\be
  \nonumber
  w(z^-,  z^+) ={\gamma^{-1} \cosh {z^- -  z^+ \ov 2} \cosh \gamma(z^- -  z^+) -  \sinh{z^- -  z^+ \ov 2} \le[ \sinh\gamma(z^- -  z^+) - \sinh v\gamma(z^- +  z^+) \ri]  \ov
\gamma^{-1} \sinh {z^- -  z^+ \ov 2} \cosh \gamma(z^- -  z^+) -  \cosh{z^- -  z^+ \ov 2} \le[ \sinh\gamma(z^- -  z^+) - \sinh v\gamma(z^- +  z^+) \ri]}
\ee
The expressions for $\tilde b$ and $\tilde w$ are too large to be presented here. A calculation yields
\be
  \nonumber
  e^\alpha = {1\ov 4e^{\beta}} = \half \le(1-{2\ov 1+v { \cosh \gamma(z^- -  z^+) \ov  \cosh v\gamma(z^- +  z^+)}} \ri)^2
\ee
and
\be
  \nonumber
  u = v= -\half .
\ee

\clearpage

\begin{figure}[h]
\begin{center}
\includegraphics[scale=0.5]{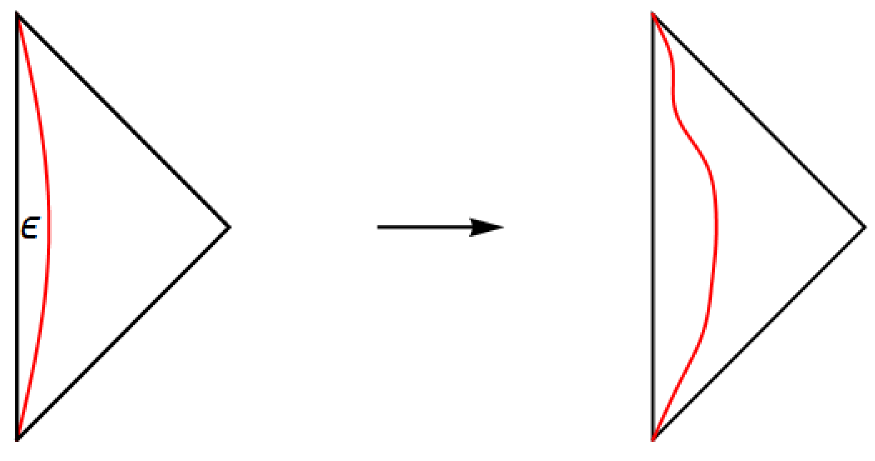} \caption{ \label{fig:conftr}
Explicitly broken conformal symmetry: AdS$_2$ is cut off a distance $\epsilon$ away from the boundary (red line). Under a conformal transformation this UV boundary moves around and therefore the value of the action changes.}

\end{center}
\end{figure}

\section{The Schwarzian  action}

\label{sec:schwarzian}

The Sachdev-Ye-Kitaev model \cite{sy, kitaev}, the 2d gravity model of Jackiw and Teitelboim  \cite{jackiw, teitelboim} and certain other gravity models \cite{Almheiri:2014cka} can be described at low energies in terms of a degree of freedom which is a function of one variable and describes reparametrizations of (imaginary) time.
The aim of this section is to derive this effective action for a string worldsheet which is an AdS$_2$ slice of the AdS$_3$ target space.

Let us therefore consider a string with a constant normal vector. The induced metric on the worldsheet is AdS$_2$. Small perturbations around the static embedding behave like a  matter field with conformal dimension\footnote{If matter is coupled to JT gravity, this dimension lies precisely at the lower end of the window $1 < \Delta < {3\ov 2}$ in which the Schwarzian action is subleading compared to the effective matter action \cite{Maldacena:2016upp}.}
$\Delta = 1$.
``Conformal symmetry'' on the boundary is the reparametrization symmetry
\be
  \label{eq:conftra}
  b \to f(b) \qquad \textrm{and} \qquad w \to f(w)
\ee
This symmetry is spontaneously broken down to $SL(2)$, i.e. simultaneous Möbius transformations
\be
  \nonumber
  b \to {c_1 + c_2 b \ov c_3 + c_4 b} \qquad \textrm{and} \qquad w \to {c_1 + c_2 w \ov c_3 + c_4 w}
\ee
which still preserve the form of the equations of motion (\ref{eq:eomwhite}) and (\ref{eq:eomblack}). There is an analogous symmetry breaking associated to the right-handed fields $\tilde b$ and $\tilde w$.

One can introduce an explicit conformal symmetry breaking by cutting out a piece of AdS$_2$ in the UV region as in FIG. \ref{fig:conftr}. Under a conformal transformation (\ref{eq:conftra}) this boundary changes and thus the value of the action also changes.
Let us use worldsheet coordinates such that $b=  z^+$ and $w= z^-$. Then the change in the action can be computed using (\ref{eq:alpha}),
\be
  \nonumber
  \Delta S =  \int dz^- dz^+  \le(e^{\alpha[f(b),f(w)]} - e^{\alpha[b,w]} \ri)
= 2  \int dz^- dz^+   \le[ {f'(z^-)f'(z^+)\ov (f(z^-)-f(z^+))^2} - {1 \ov (z^- - z^+)^2} \ri]
\ee
In FIG. \ref{fig:conftr}, the original AdS$_2$ boundary is the vertical black line. On the boundary, retarded and advanced times are equal, i.e. $b=w$. In our gauge this is at $z^- =  z^+$. Let us now introduce the UV cutoff (red line in the figure) at
\be
  \nonumber
  z^- = z^-_\epsilon \equiv  z^+ + \epsilon \varphi( z^+)
\ee
Here we have included a varying coupling $\varphi( z^+)$.
After performing the $z^-$ integral between $-\infty$ and the UV cutoff $z_\epsilon^-$, we get
\be
  \nonumber
  \Delta S = 2 \int dz^+  \le[ {f'(z^+)\ov f(z^+)-f(z_\epsilon^-)} - {1 \ov  z^+ -z_\epsilon^-} \ri]
\ee
For small $\epsilon$, the integrand can be Taylor expanded. Assuming $f'(+\infty) = f'(-\infty)$ we get
\be
  \nonumber
  \Delta S \approx {2 \epsilon \ov 3} \int dz^+ \,  \varphi( z^+)  \le[ {f'''(z^+) \ov f'(z^+)} - {3\ov 2} {f''(z^+)^2 \ov f'(z^+)^2}  \ri]
\ee
This is the Schwarzian action \cite{kitaev, Maldacena:2016hyu, Jensen:2016pah, Maldacena:2016upp, Engelsoy:2016xyb}.
Related expressions for the renormalized string area in the Euclidean case were obtained in \cite{Kruczenski:2014bla} (see also \cite{Irrgang:2015txa, Dekel:2015bla, Huang:2016atz, He:2017cwd}).
In the Lorentzian case the above Schwarzian action for the string was computed in \cite{Banerjee:2018twd}.

\subsection{An exact Schwarzian on the worldsheet}

The Euler-Lagrange equations for the Schwarzian action produce a fourth order differential equation. Note that if  $w$ is expressed using (\ref{eq:eomblack}) as
\be
  \nonumber
  w = b- {2 \p_- b  \,  \p_+ b \over \p_- \p_+ b }
\ee
then  (\ref{eq:eomwhite}) will give a fourth order equation for $b$.

Let us compute the change in the action if $w$ is kept on-shell and $b$ is transformed  $b \to f(b)$.
We obtain
\be
  \nonumber
  \Delta S =  \int dz^- dz^+  \le(e^{\alpha[f(b)]} - e^{\alpha[b]} \ri)
= \half  \int dz^- dz^+   \le[{f'''(b) \ov f'(b)} - {3\ov 2} {f''(b)^2 \ov f'(b)^2} \ri]  \p_- b \, \p_+ b
\ee
It is interesting to see the appearance of the Schwarzian again. In this case it is integrated over the two-dimensional worldsheet. Note, however, that this expression for the action is exact and it does not require the embedding to be special (e.g. close to an AdS$_2$ slice).

\clearpage

\section{Discussion}

In this paper we have considered a Nambu-Goto string in AdS$_3$ spacetime. We have mainly used the embedding into ambient space: $Y(z^-, z^+) \in $ AdS$_3$ $ \subset \RR^{2,2}$ (here $z^\pm$ are null coordinates on the Lorentzian worldsheet).
By taking a continuum limit of the discrete equation of motion of segmented strings, we have obtained differential equations for a smooth string in terms of two fields
\be
  \nonumber
  b(z^-, z^+) \equiv \mathfrak{a}(\p_- Y)
  \qquad
  w(z^-, z^+) \equiv \mathfrak{a}( \p_+ Y)
\ee
where $\mathfrak{a}(X) = {X_{-1} + X_2 \over X_0 + X_1}$. Our main results are the following equations of motion\footnote{Note that these equations are the Lorentzian analog of those coming from a torus compactification, i.e.
\be
  \nonumber
  \p \bar \p \tau + {2 \p \tau \bar \p \tau \ov \bar \tau - \tau } = 0
\ee
where $\tau(z,\bar z)\in \CC$ is the complex structure parameter of the torus fiber and $z = x_1 + i x_2$ is the two-dimensional Euclidean base \cite{Greene:1989ya}. In the case of the string in AdS, the base is the Lorentzian worldsheet which enjoys a conformal symmetry. Furthermore, $\tau_2$ has to be Wick-rotated to purely imaginary values.}
\be
  \nonumber
  \p_-  \p_+ w = 2{ \p_- w \,  \p_+ w \over w-b } , \qquad
\p_-  \p_+ b = 2{ \p_- b \,  \p_+ b \over b-w } .
\ee
We define the following auxiliary fields: the area density $e^{\alpha}$, the area density in the space of normal vectors $e^{\beta}$, and the  $u$ and $v$ fields. They are given by
\be
  \nonumber
  e^{\alpha(z^-, z^+)} = 2{\p_+ b \, \p_- w  \over (b-w)^2} , \qquad
  e^{\beta(z^-, z^+)} =  2{ \p_- b \,  \p_+ w \over (b-w)^2} ,
\ee
\be
  \nonumber
  u(z^-) =  2{ \p_- b \, \p_- w \over (b-w)^2} , \qquad
  v(z^+) =  2{ \p_+ b \, \p_+ w \over (b-w)^2} .
\ee
Using the equations of motion it is easy to see that the auxiliary fields satisfy the generalized sinh-Gordon equations \cite{Pohlmeyer:1975nb}
\be
  \nonumber
   \p_- \p_+ \alpha + e^{\alpha} - uv  e^{-\alpha} = 0 ,
\ee
\be
  \nonumber
   \p_- \p_+ \beta + e^{\beta} - uv  e^{-\beta} = 0 .
\ee

There are analogous equations for the right-handed fields that we denote with a tilde. Instead of $\mathfrak{a}$, these are defined using $\mathfrak{\tilde a}( X)  = {X_{-1} + X_2 \over -X_0 + X_1}$.
The right-handed fields are not independent, because the constraints
\be
  \nonumber
  \alpha =  {\tilde \alpha } , \qquad    \beta =  {\tilde \beta } , \qquad
  u = -\tilde u , \qquad v = -\tilde v
\ee
must be satisfied by the $b$, $w$, $\tilde b$, $\tilde w$ fields. In particular, $\alpha =  {\tilde \alpha }$ means that the right-handed variables must give the same induced metric on the worldsheet as the left-handed ones.

\clearpage

A potential application of the celestial variables developed in \cite{Vegh:2016hwq, Vegh:2016fcm} and in the current paper is an alternative calculation of the spectral curve. Preliminary results show that in certain cases it is possible to write down explicit expressions for the quasimomenta without the use of transcendental functions \cite{vegh1}.

The theory on the string worldsheet shares many features with theories of quantum gravity \cite{Dubovsky:2012wk}.
The $b$ and $w$ fields are in a certain sense {\it holographic}: they describe the string embedding using coordinates  at null infinity. One might hope that the string worldsheet has a holographic description and that the celestial fields play a role in this.




In the flat space (or the dual near-AdS$_2$) limit the generalized sinh-Gordon equation degenerates and becomes the Liouville equation. This can be solved explicitly in terms of two arbitrary function (see section \ref{sec:flat}).
As one can see from the expression for $e^{\alpha}$ in (\ref{eq:alpha}), the celestial fields provide a generalization of the Liouville solution which is applicable to the sinh-Gordon equation. Although the description contains two fields, there is only one physical degree of freedom: the transverse position of the string in the AdS$_3$ target space.

The string embedding can be computed from a sinh-Gordon solution if one solves an associated Dirac scattering problem, see e.g. \cite{Alday:2009yn}. 
In the Appendix we discuss how the black and white celestial fields can be expressed as ratios of the spinor components of the solution.


In this paper we have used \poincare coordinates. In global AdS coordinates the form of the equation change. This can be seen by replacing $b \to \tan b$ and $w \to \tan w$. The new celestial fields now  correspond to retarded and advanced boundary times in {\it global} AdS coordinates at which null rays emanating from the string reach the boundary. Due to the nature of the immersion of AdS$_3$ into $\RR^{2,2}$ in (\ref{eq:hyp}) the time coordinates are periodic, therefore $b$ and $w$ are coordinates on a torus. By plugging in the tangent functions into (\ref{eq:eomwhite}) and (\ref{eq:eomblack}) it is easy to see that they satisfy the equations
\be
  \nonumber
  \p_-  \p_+ w = 2{ \p_- w \,  \p_+ w \over \tan(w-b) } , \qquad
\p_-  \p_+ b = 2{ \p_- b \,  \p_+ b \over \tan(b-w) } .
\ee    
Similar equations are also expected to be valid when the target space is de Sitter spacetime \cite{Pohlmeyer:1975nb, Gubser:2016wno}. The string worldsheet can also be coupled to a background two-form whose field strength is the volume form of AdS$_3$ (i.e. the coupling does not destroy any of the continuous symmetries of the system \cite{Gubser:2016zyw}). Independently of the value of the coupling, the discrete equation of motion in (\ref{eq:deqn}) remains the same \cite{Vegh:2016fcm} and thus the $b$ and $w$ fields will satisfy the same equations of motion. Finally, a higher-dimensional generalization may be possible since the spinor-helicity formalism has been extended to higher dimensions (see e.g \cite{Cheung:2009dc}). In AdS$_3$  left- and right-handed celestial fields ($b, w$ and $\tilde b, \tilde w$, respectively) decouple. Thus one can concentrate on either pair without having to deal with constraints between them.  In general dimensions one does not expect such  a decoupling. This makes AdS$_3$ special.

\vspace{0.2in}   \centerline{\bf{Acknowledgments}} \vspace{0.2in}
I thank Martin Kruczenski, Douglas Stanford, and the referee for valuable comments on the manuscript.
The author is supported by the STFC Ernest Rutherford grant ST/P004334/1.


\section*{Appendix}

The string embedding can be computed from a sinh-Gordon solution if one solves an associated scattering problem of a spinor field. In this Appendix we will discuss how the black and white celestial fields can be expressed as ratios of the spinor components of the solution.
The $SL(2)$ Lax matrices are  \cite{Alday:2009yn}
\bea
\nonumber
B^-_L=\begin{pmatrix}
{1 \over 4} \p_- \alpha & {1\ov \sqrt{2}} {e^{ {\alpha \ov 2}}  }\cr
{u\ov \sqrt{2}}{e^{-  {\alpha \ov 2}}  }  & -{1 \over 4} \p_- \alpha
\end{pmatrix} &&
B^+_L=\begin{pmatrix}
-{1 \over 4} \p_+ \alpha & -{v \ov \sqrt{2}}{e^{-{\alpha \ov 2} }   }\cr
-{1\ov \sqrt{2}} e^{ {\alpha \ov 2}}    & {1 \over 4} \p_+ \alpha
\end{pmatrix} \\
\nonumber
B^-_R=\begin{pmatrix}
-{1 \over 4} \p_- \alpha &{-{u\ov \sqrt{2}} e^{- {\alpha \ov 2}}   }   \cr
{1\ov \sqrt{2}}{e^{ {\alpha \ov 2}}  } & {1 \over 4} \p_- \alpha
\end{pmatrix}
&&
B^+_R=\begin{pmatrix}{1 \over 4} \p_+ \alpha & -{1\ov \sqrt{2}}{e^{ {\alpha \ov 2}} }  \cr
{v\ov \sqrt{2}}{e^{- {\alpha \ov 2}}   }  & -{1 \over 4} \p_+ \alpha \end{pmatrix} \, .
\eea
The flatness conditions
\bea
\nonumber
\p_- B^+_L-\p_+ B^-_L+[B^-_L,B^+_L] &= & 0 \, , \\
\nonumber
\p_- B^+_R-\p_+ B^-_R+[B^-_R,B^+_R] &=& 0
\eea
are   equivalent to the  sinh-Gordon equation.
Let us now consider the left and right   linear systems
\bea
\nonumber
\p_- \psi^L_{\alpha}+(B^-_L)_{\alpha}^{~\beta}\psi^L_{\beta}=0 \, , & &
\p_+ \psi^L_{\alpha}+(B^+_L)_{\alpha}^{~\beta}\psi^L_{\beta}=0 \, , \\
\label{eq:linear}
\p_- \psi^R_{\dot \alpha }+(B^-_R)_{\dot \alpha}^{~\dot \beta}\psi^R_{\dot \beta } =0 \, , & &
\p_+ \psi^R_{\dot \alpha }+(B^+_R)_{\dot \alpha}^{~\dot \beta}\psi^R_{\dot \beta } =0 \, . \quad
\eea
where $\psi^L_{\alpha}$ and $\psi^R_{\dot \alpha }$ are two-component spinors ($\alpha, \dot \alpha \in \{ 0,1 \}$). Each of these systems have two linearly independent solutions, denoted by $\psi^L_{\alpha a}$   and $\psi^R_{\dot \alpha \dot a}$ (here $ a, \dot  a \in \{ 0,1 \}$ label the independent solutions).
We will normalize them so that
\be
  \epsilon^{\beta \alpha } \psi^L_{\alpha a} \psi^L_{\beta b}=\epsilon_{a b},~~~~~\epsilon^{\dot \beta \dot \alpha } \psi^R_{\dot \alpha  \dot a} \psi^R_{\dot \beta \dot b}=\epsilon_{\dot a \dot b}
\ee
where $\epsilon$ is the $2\times 2$ Levi-Civita tensor. Using the spinor solutions, one can compute the matrix  \cite{Alday:2009yn}
\be
  \nonumber
  W= \half \begin{pmatrix}
Y+N & {e^{ -{\alpha \ov 2 }}\p_+ Y  }\cr
 {e^{-  {\alpha\ov 2 }}  }\p_- Y  & Y-N
\end{pmatrix} \, ,
\ee
where $Y(z^-,  z^+)$ is the embedding function, $N(z^-,  z^+)$ is the normal vector function. Let $\alpha, \dot \alpha$ denote the indices of the $2\times 2$ matrix, and use the equivalence of $SO(2,2)$ and $SL(2)\times SL(2)$ to decompose spacetime indices into spinor indices $a, \dot a$. The result is a  quantity with four spinor indices $W = W_{\alpha\dot \alpha, a\dot a}$.
The trace computes the string embedding which is given by 
\be
 Y_{a \dot a } \equiv
\begin{pmatrix}
Y_{-1}+Y_{2}& Y_1-Y_0\cr Y_1 + Y_0 & Y_{-1}-Y_{2}
\end{pmatrix}_{a,\dot{a}}=
\psi^L_{\alpha,a} M_1^{\alpha \dot \beta} \psi^R_{\dot \beta ,\dot{a}}
\ee
where $M_1 = \textrm{diag}(1,1)$.
Similarly, one can compute  $e^{-{\alpha\ov 2 }}\p_\pm Y$ by replacing $M_1$ with another matrix. By taking ratios one can obtain the $b$ and $w$ fields, 
\be
  \nonumber
 b(z^-,  z^+)  = {(\p_- Y)_{-1}+ (\p_- Y)_{2} \over  (\p_- Y)_{1}+(\p_- Y)_{0}} = {\psi^L_{10} \psi^R_{\dot{0}\dot{0}}  \ov \psi^L_{11} \psi^R_{\dot{0}\dot{0}} }= {\psi^L_{10}    \ov \psi^L_{11}  } \, .
\ee
Similarly, we have
\be
  \nonumber
    w(z^-,  z^+)  = {(\p_+ Y)_{-1}+ (\p_+ Y)_{2}  \over (\p_+ Y)_{0}+ (\p_+ Y)_{1}} =  {\psi^L_{00}    \ov \psi^L_{01}  } \, .
\ee
Note that the right spinors have dropped out from the final expressions.

One can also compute the `right-handed' celestial fields
\be
  \nonumber
 \tilde b(z^-,  z^+)  = {(\p_- Y)_{-1}+ (\p_- Y)_{2} \over  (\p_- Y)_{1}-(\p_- Y)_{0}} = {\psi^L_{10} \psi^R_{\dot{0}\dot{0}}  \ov \psi^L_{10} \psi^R_{\dot{0}\dot{1}} } = { \psi^R_{\dot{0}\dot{0}}  \ov  \psi^R_{\dot{0}\dot{1}} }  \, ,
\ee
and
\be
  \nonumber
   \tilde w(z^-,  z^+)  = {(\p_+ Y)_{-1}+ (\p_+ Y)_{2}  \over (\p_+ Y)_{1}- (\p_+ Y)_{0}} =  {\psi^R_{\dot 1 \dot 0}    \ov \psi^R_{\dot 1 \dot 1}  } \, .
\ee
These formulas relate the celestial fields to the solutions of the linear problem.

\clearpage

\bibliographystyle{JHEP}
\bibliography{paper}

\providecommand{\href}[2]{#2}\begingroup\raggedright\begin{thebibliography}{10}

\bibitem{Ishii:2015wua}
T.~Ishii and K.~Murata, {\it {Turbulent strings in AdS/CFT}},  {\em JHEP} {\bf
  06} (2015) 086 [\href{http://arXiv.org/abs/1504.02190}{{\tt 1504.02190}}].

\bibitem{Vegh:2018dda}
D.~Vegh, {\it {Pair-production of cusps on a string in AdS$_3$}},
  \href{http://arXiv.org/abs/1802.04306}{{\tt 1802.04306}}.

\bibitem{Maldacena:1997re}
J.~M. Maldacena, {\it {The large N limit of superconformal field theories and
  supergravity}},  {\em Adv. Theor. Math. Phys.} {\bf 2} (1998) 231--252.

\bibitem{Gubser:1998bc}
S.~S. Gubser, I.~R. Klebanov and A.~M. Polyakov, {\it {Gauge theory correlators
  from non-critical string theory}},  {\em Phys. Lett.} {\bf B428} (1998)
  105--114 [\href{http://arXiv.org/abs/hep-th/9802109}{{\tt hep-th/9802109}}].

\bibitem{Witten:1998qj}
E.~Witten, {\it {Anti-de Sitter space and holography}},  {\em Adv. Theor. Math.
  Phys.} {\bf 2} (1998) 253--291.

\bibitem{Herzog:2006gh}
C.~P. Herzog, A.~Karch, P.~Kovtun, C.~Kozcaz and L.~G. Yaffe, {\it {Energy loss
  of a heavy quark moving through N=4 supersymmetric Yang-Mills plasma}},  {\em
  JHEP} {\bf 07} (2006) 013 [\href{http://arXiv.org/abs/hep-th/0605158}{{\tt
  hep-th/0605158}}].

\bibitem{Liu:2006ug}
H.~Liu, K.~Rajagopal and U.~A. Wiedemann, {\it {Calculating the jet quenching
  parameter from AdS/CFT}},  {\em Phys. Rev. Lett.} {\bf 97} (2006) 182301
  [\href{http://arXiv.org/abs/hep-ph/0605178}{{\tt hep-ph/0605178}}].

\bibitem{Gubser:2006bz}
S.~S. Gubser, {\it {Drag force in AdS/CFT}},  {\em Phys. Rev.} {\bf D74} (2006)
  126005 [\href{http://arXiv.org/abs/hep-th/0605182}{{\tt hep-th/0605182}}].

\bibitem{Maldacena:2015waa}
J.~Maldacena, S.~H. Shenker and D.~Stanford, {\it {A bound on chaos}},  {\em
  JHEP} {\bf 08} (2016) 106 [\href{http://arXiv.org/abs/1503.01409}{{\tt
  1503.01409}}].

\bibitem{Shenker:2013pqa}
S.~H. Shenker and D.~Stanford, {\it {Black holes and the butterfly effect}},
  {\em JHEP} {\bf 03} (2014) 067 [\href{http://arXiv.org/abs/1306.0622}{{\tt
  1306.0622}}].

\bibitem{Shenker:2013yza}
S.~H. Shenker and D.~Stanford, {\it {Multiple Shocks}},  {\em JHEP} {\bf 12}
  (2014) 046 [\href{http://arXiv.org/abs/1312.3296}{{\tt 1312.3296}}].

\bibitem{Murata:2017rbp}
K.~Murata, {\it {Fast scrambling in holographic Einstein-Podolsky-Rosen pair}},
   {\em JHEP} {\bf 11} (2017) 049 [\href{http://arXiv.org/abs/1708.09493}{{\tt
  1708.09493}}].

\bibitem{deBoer:2017xdk}
J.~de~Boer, E.~Llabres, J.~F. Pedraza and D.~Vegh, {\it {Chaotic strings in
  AdS/CFT}},  {\em Phys. Rev. Lett.} {\bf 120} (2018), no.~20 201604
  [\href{http://arXiv.org/abs/1709.01052}{{\tt 1709.01052}}].

\bibitem{Dubovsky:2012wk}
S.~Dubovsky, R.~Flauger and V.~Gorbenko, {\it {Solving the Simplest Theory of
  Quantum Gravity}},  {\em JHEP} {\bf 09} (2012) 133
  [\href{http://arXiv.org/abs/1205.6805}{{\tt 1205.6805}}].

\bibitem{Pohlmeyer:1975nb}
K.~Pohlmeyer, {\it {Integrable Hamiltonian Systems and Interactions Through
  Quadratic Constraints}},  {\em Commun. Math. Phys.} {\bf 46} (1976) 207--221.

\bibitem{DeVega:1992xc}
H.~J. De~Vega and N.~G. Sanchez, {\it {Exact integrability of strings in
  D-Dimensional De Sitter space-time}},  {\em Phys. Rev.} {\bf D47} (1993)
  3394--3405.

\bibitem{Bena:2003wd}
I.~Bena, J.~Polchinski and R.~Roiban, {\it {Hidden symmetries of the AdS(5) x
  S**5 superstring}},  {\em Phys. Rev.} {\bf D69} (2004) 046002
  [\href{http://arXiv.org/abs/hep-th/0305116}{{\tt hep-th/0305116}}].

\bibitem{Beisert:2010jr}
N.~Beisert {\em et.~al.}, {\it {Review of AdS/CFT Integrability: An Overview}},
   {\em Lett. Math. Phys.} {\bf 99} (2012) 3--32
  [\href{http://arXiv.org/abs/1012.3982}{{\tt 1012.3982}}].

\bibitem{Tseytlin:2010jv}
A.~A. Tseytlin, {\it {Review of AdS/CFT Integrability, Chapter II.1: Classical
  AdS5xS5 string solutions}},  {\em Lett. Math. Phys.} {\bf 99} (2012) 103--125
  [\href{http://arXiv.org/abs/1012.3986}{{\tt 1012.3986}}].

\bibitem{Gubser:2002tv}
S.~S. Gubser, I.~R. Klebanov and A.~M. Polyakov, {\it {A Semiclassical limit of
  the gauge / string correspondence}},  {\em Nucl. Phys.} {\bf B636} (2002)
  99--114 [\href{http://arXiv.org/abs/hep-th/0204051}{{\tt hep-th/0204051}}].

\bibitem{Kruczenski:2004wg}
M.~Kruczenski, {\it {Spiky strings and single trace operators in gauge
  theories}},  {\em JHEP} {\bf 08} (2005) 014
  [\href{http://arXiv.org/abs/hep-th/0410226}{{\tt hep-th/0410226}}].

\bibitem{Kalousios:2006xy}
C.~Kalousios, M.~Spradlin and A.~Volovich, {\it {Dressing the giant magnon
  II}},  {\em JHEP} {\bf 03} (2007) 020
  [\href{http://arXiv.org/abs/hep-th/0611033}{{\tt hep-th/0611033}}].

\bibitem{Alday:2007hr}
L.~F. Alday and J.~M. Maldacena, {\it {Gluon scattering amplitudes at strong
  coupling}},  {\em JHEP} {\bf 06} (2007) 064
  [\href{http://arXiv.org/abs/0705.0303}{{\tt 0705.0303}}].

\bibitem{Grigoriev:2007bu}
M.~Grigoriev and A.~A. Tseytlin, {\it {Pohlmeyer reduction of AdS(5) x S**5
  superstring sigma model}},  {\em Nucl. Phys.} {\bf B800} (2008) 450--501
  [\href{http://arXiv.org/abs/0711.0155}{{\tt 0711.0155}}].

\bibitem{Jevicki:2007aa}
A.~Jevicki, K.~Jin, C.~Kalousios and A.~Volovich, {\it {Generating AdS String
  Solutions}},  {\em JHEP} {\bf 0803} (2008) 032
  [\href{http://arXiv.org/abs/0712.1193}{{\tt 0712.1193}}].

\bibitem{Jevicki:2008mm}
A.~Jevicki and K.~Jin, {\it {Solitons and AdS String Solutions}},  {\em Int. J.
  Mod. Phys.} {\bf A23} (2008) 2289--2298
  [\href{http://arXiv.org/abs/0804.0412}{{\tt 0804.0412}}].

\bibitem{Dorey:2008vp}
N.~Dorey and M.~Losi, {\it {Spiky Strings and Spin Chains}},
  \href{http://arXiv.org/abs/0812.1704}{{\tt 0812.1704}}.

\bibitem{Jevicki:2009uz}
A.~Jevicki and K.~Jin, {\it {Moduli Dynamics of AdS(3) Strings}},  {\em JHEP}
  {\bf 06} (2009) 064 [\href{http://arXiv.org/abs/0903.3389}{{\tt 0903.3389}}].

\bibitem{Dorey:2010iy}
N.~Dorey and M.~Losi, {\it {Giant Holes}},  {\em J. Phys.} {\bf A43} (2010)
  285402 [\href{http://arXiv.org/abs/1001.4750}{{\tt 1001.4750}}].

\bibitem{Dorey:2010id}
N.~Dorey and M.~Losi, {\it {Spiky Strings and Giant Holes}},  {\em JHEP} {\bf
  12} (2010) 014 [\href{http://arXiv.org/abs/1008.5096}{{\tt 1008.5096}}].

\bibitem{Irrgang:2012xb}
A.~Irrgang and M.~Kruczenski, {\it {Rotating Wilson loops and open strings in
  AdS3}},  {\em J.Phys.} {\bf A46} (2013) 075401
  [\href{http://arXiv.org/abs/1210.2298}{{\tt 1210.2298}}].

\bibitem{Vegh:2015ska}
D.~Vegh, {\it {The broken string in anti-de Sitter space}},  {\em JHEP} {\bf
  02} (2018) 045 [\href{http://arXiv.org/abs/1508.06637}{{\tt 1508.06637}}].

\bibitem{Callebaut:2015fsa}
N.~Callebaut, S.~S. Gubser, A.~Samberg and C.~Toldo, {\it {Segmented strings in
  AdS$_{3}$}},  {\em JHEP} {\bf 11} (2015) 110
  [\href{http://arXiv.org/abs/1508.07311}{{\tt 1508.07311}}].

\bibitem{Vegh:2015yua}
D.~Vegh, {\it {Colliding waves on a string in AdS$_3$}},
  \href{http://arXiv.org/abs/1509.05033}{{\tt 1509.05033}}.

\bibitem{Gubser:2016wno}
S.~S. Gubser, {\it {Evolution of segmented strings}},  {\em Phys. Rev.} {\bf
  D94} (2016), no.~10 106007 [\href{http://arXiv.org/abs/1601.08209}{{\tt
  1601.08209}}].

\bibitem{Gubser:2016zyw}
S.~S. Gubser, S.~Parikh and P.~Witaszczyk, {\it {Segmented strings and the
  McMillan map}},  {\em JHEP} {\bf 07} (2016) 122
  [\href{http://arXiv.org/abs/1602.00679}{{\tt 1602.00679}}].

\bibitem{Vegh:2016hwq}
D.~Vegh, {\it {Segmented strings from a different angle}},
  \href{http://arXiv.org/abs/1601.07571}{{\tt 1601.07571}}.

\bibitem{Vegh:2016fcm}
D.~Vegh, {\it {Segmented strings coupled to a B-field}},  {\em JHEP} {\bf 04}
  (2018) 088 [\href{http://arXiv.org/abs/1603.04504}{{\tt 1603.04504}}].

\bibitem{Artru:1979ye}
X.~Artru, {\it {Classical String Phenomenology. 1. How Strings Work}},  {\em
  Phys. Rept.} {\bf 97} (1983) 147.

\bibitem{Andersson:1983ia}
B.~Andersson, G.~Gustafson, G.~Ingelman and T.~Sjostrand, {\it {Parton
  Fragmentation and String Dynamics}},  {\em Phys. Rept.} {\bf 97} (1983)
  31--145.

\bibitem{suris}
Y.~B. Suris, {\em The Problem of Integrable Discretization: Hamiltonian
  Approach}.
\newblock Birkh\" auser Verlag, Basel, 2003.

\bibitem{Pasterski:2016qvg}
S.~Pasterski, S.-H. Shao and A.~Strominger, {\it {Flat Space Amplitudes and
  Conformal Symmetry of the Celestial Sphere}},  {\em Phys. Rev. D} {\bf 96}
  (2017), no.~6 065026 [\href{http://arXiv.org/abs/1701.00049}{{\tt
  1701.00049}}].

\bibitem{Atanasov:2021oyu}
A.~Atanasov, A.~Ball, W.~Melton, A.-M. Raclariu and A.~Strominger, {\it
  {$(2,2)$ Scattering and the Celestial Torus}},
  \href{http://arXiv.org/abs/2101.09591}{{\tt 2101.09591}}.

\bibitem{Kruczenski:2014bla}
M.~Kruczenski, {\it {Wilson loops and minimal area surfaces in hyperbolic
  space}},  {\em JHEP} {\bf 11} (2014) 065
  [\href{http://arXiv.org/abs/1406.4945}{{\tt 1406.4945}}].

\bibitem{Mikhailov:2003er}
A.~Mikhailov, {\it {Nonlinear waves in AdS / CFT correspondence}},
  \href{http://arXiv.org/abs/hep-th/0305196}{{\tt hep-th/0305196}}.

\bibitem{sy}
S.~Sachdev and J.~Ye, {\it Gapless spin-fluid ground state in a random quantum
  heisenberg magnet},  {\em Phys. Rev. Lett.} {\bf 70} (May, 1993) 3339--3342.

\bibitem{kitaev}
A.~Kitaev, {\it {A Simple Model of Quantum Holography, Talks at KITP, April 7,
  2015 and May 27, 2015, http://online.kitp.ucsb.edu/online/entangled15/kitaev
  }}, .

\bibitem{jackiw}
R.~Jackiw, {\it Lower dimensional gravity},  {\em Nuclear Physics B} {\bf 252}
  (1985) 343 -- 356.

\bibitem{teitelboim}
C.~Teitelboim, {\it Gravitation and hamiltonian structure in two spacetime
  dimensions},  {\em Physics Letters B} {\bf 126} (1983), no.~1 41 -- 45.

\bibitem{Almheiri:2014cka}
A.~Almheiri and J.~Polchinski, {\it {Models of AdS$_{2}$ backreaction and
  holography}},  {\em JHEP} {\bf 11} (2015) 014
  [\href{http://arXiv.org/abs/1402.6334}{{\tt 1402.6334}}].

\bibitem{Maldacena:2016upp}
J.~Maldacena, D.~Stanford and Z.~Yang, {\it {Conformal symmetry and its
  breaking in two dimensional Nearly Anti-de-Sitter space}},  {\em PTEP} {\bf
  2016} (2016), no.~12 12C104 [\href{http://arXiv.org/abs/1606.01857}{{\tt
  1606.01857}}].

\bibitem{Maldacena:2016hyu}
J.~Maldacena and D.~Stanford, {\it {Remarks on the Sachdev-Ye-Kitaev model}},
  {\em Phys. Rev.} {\bf D94} (2016), no.~10 106002
  [\href{http://arXiv.org/abs/1604.07818}{{\tt 1604.07818}}].

\bibitem{Jensen:2016pah}
K.~Jensen, {\it {Chaos in AdS$_2$ Holography}},  {\em Phys. Rev. Lett.} {\bf
  117} (2016), no.~11 111601 [\href{http://arXiv.org/abs/1605.06098}{{\tt
  1605.06098}}].

\bibitem{Engelsoy:2016xyb}
J.~Engelsoy, T.~G. Mertens and H.~Verlinde, {\it {An investigation of AdS2
  backreaction and holography}},  {\em JHEP} {\bf 07} (2016) 139
  [\href{http://arXiv.org/abs/1606.03438}{{\tt 1606.03438}}].

\bibitem{Irrgang:2015txa}
A.~Irrgang and M.~Kruczenski, {\it {Euclidean Wilson loops and minimal area
  surfaces in lorentzian AdS3}},  {\em JHEP} {\bf 12} (2015) 083
  [\href{http://arXiv.org/abs/1507.02787}{{\tt 1507.02787}}].

\bibitem{Dekel:2015bla}
A.~Dekel, {\it {Wilson Loops and Minimal Surfaces Beyond the Wavy
  Approximation}},  {\em JHEP} {\bf 03} (2015) 085
  [\href{http://arXiv.org/abs/1501.04202}{{\tt 1501.04202}}].

\bibitem{Huang:2016atz}
C.~Huang, Y.~He and M.~Kruczenski, {\it {Minimal area surfaces dual to Wilson
  loops and the Mathieu equation}},  {\em JHEP} {\bf 08} (2016) 088
  [\href{http://arXiv.org/abs/1604.00078}{{\tt 1604.00078}}].

\bibitem{He:2017cwd}
Y.~He, C.~Huang and M.~Kruczenski, {\it {Minimal area surfaces in AdS(n+1) and
  Wilson loops}},  {\em JHEP} {\bf 02} (2018) 027
  [\href{http://arXiv.org/abs/1712.06269}{{\tt 1712.06269}}].

\bibitem{Banerjee:2018twd}
A.~Banerjee, A.~Kundu and R.~R. Poojary, {\it {Strings, Branes, Schwarzian
  Action and Maximal Chaos}},  \href{http://arXiv.org/abs/1809.02090}{{\tt
  1809.02090}}.

\bibitem{Greene:1989ya}
B.~R. Greene, A.~D. Shapere, C.~Vafa and S.-T. Yau, {\it Stringy cosmic strings
  and noncompact {C}alabi-{Y}au manifolds},  {\em Nucl. Phys.} {\bf B337}
  (1990) 1.

\bibitem{vegh1}
D.~Vegh, ``Work in progress.''

\bibitem{Alday:2009yn}
L.~F. Alday and J.~Maldacena, {\it {Null polygonal Wilson loops and minimal
  surfaces in Anti-de-Sitter space}},  {\em JHEP} {\bf 0911} (2009) 082
  [\href{http://arXiv.org/abs/0904.0663}{{\tt 0904.0663}}].

\bibitem{Cheung:2009dc}
C.~Cheung and D.~O'Connell, {\it {Amplitudes and Spinor-Helicity in Six
  Dimensions}},  {\em JHEP} {\bf 07} (2009) 075
  [\href{http://arXiv.org/abs/0902.0981}{{\tt 0902.0981}}].

\end{thebibliography}\endgroup

\end{document}